\documentstyle[aps,amssymb,eqsecnum,tighten, epsf]{revtex}
\begin{document}
\title{Moving Black Holes in 3D}
\author{R. G\'{o}mez,
	L. Lehner, 
	R. L. Marsa, and  
	J. Winicour}
\address{
	 Department of Physics and Astronomy,
 	 University of Pittsburgh,
 	 Pittsburgh, PA 15260 }
\maketitle

\begin{abstract}

We model a radiating, moving black hole in terms of a worldtube-nullcone
boundary value problem. We evolve this data in the region interior to
the worldtube but exterior to a trapped surface by means of a
characteristic evolution based upon a family of ingoing null
hypersurfaces. Data on the worldtube is induced from a Schwarzschild
spacetime but the worldtube is allowed to move relative to the static
Schwarzschild trajectories. When the worldtube is stationary (static or
rotating in place), a distorted black hole inside it evolves to
equilibrium with the Schwarzschild boundary. A boost of the worldtube
with respect to the Schwarzschild black hole does not affect these
results. The code also stably tracks an unlimited number of orbits when
the worldtube wobbles periodically. The work establishes that
characteristic evolution can evolve a spacetime with a distorted black hole
moving on a 3-dimensional grid with the controlled accuracy and long
term stability necessary to investigate new facets of black hole
physics.

\end{abstract}

\section{Introduction}

The calculation of gravitational waveforms from the inspiral and merger
of binary black holes is a great current challenge to computational
relativity with important relevance to LIGO. A three dimensional
computer code is now being constructed by the Binary Black Hole Grand
Challenge Alliance to solve this problem by Cauchy evolution of initial
data for two black holes \cite{science}. The most difficult problems
with the development of this code are inaccuracy and instability at the
boundaries of the Cauchy grid. Matching to an exterior characteristic
evolution is one method being pursued to handle the outer Cauchy
boundary and to extract the waveform at null infinity.  This has been
shown to be a highly accurate and efficient approach in the treatment
of 3-dimensional nonlinear scalar waves \cite{bishop1996,jcp}. For the
purpose of extending this approach to full general relativity, a
3-dimensional characteristic code for the gravitational field has been
developed and fully calibrated to perform with second order accuracy
and robust stability in highly nonlinear regimes \cite{cce,highp}. A
module for matching Cauchy and characteristic gravitational evolution
codes across a worldtube interface has also been written \cite{manual}
and is now in the testing stage. In this paper, we present results
which show that Cauchy-characteristic matching can also solve the very
difficult inner boundary condition necessary for Cauchy evolution of
black holes. Notably, we have used a characteristic code to achieve the
first successful treatment of a distorted black hole moving on a
3-dimensional grid with apparently unlimited long term stability.

The conventional strategy for avoiding the topological and strong field
difficulties in the Cauchy evolution of black holes has been to excise
the region interior to an apparent horizon, as initially suggested by
W. Unruh \cite{thornburg1987}. In a recent work \cite{excise}, we
proposed an alternative version of this strategy in which the black
hole region is evolved by a characteristic evolution based upon ingoing
null cones. These null cones are truncated at an inner boundary
consisting of a trapped or marginally trapped surface, and matched at
their outer boundary to the inner boundary of a Cauchy evolution. In
turn, the outer boundary of the Cauchy evolution is matched to an
exterior characteristic evolution extending to (compactified) infinity.
The potential advantages over a purely Cauchy approach to the inner
boundary were discussed in that work and the global strategy was
successfully implemented for spherically symmetric self-gravitating
scalar waves evolving in a black hole spacetime.

We present here the successful implementation of a characteristic
treatment of an inner black hole boundary for fully 3-dimensional
simulations containing gravitational radiation. We show that the
ingoing characteristic approach is able to locate the black hole and to
track it stably as it moves on the numerical grid. For a report on
recent progress in tackling the same problem by Cauchy evolution see
Ref.~\cite{losalam}. The boundary data for the characteristic initial
value problem is posed on a worldtube and on an ingoing null cone
emanating from the initial slice of the worldtube. A main goal of this
study is to develop new methods which will allow a combination of
characteristic and Cauchy evolution to tackle the computational problem
of the inspiral of two black holes. In this new approach, the two
evolutions are matched across a worldtube, with the Cauchy domain
supplying the boundary values for characteristic evolution and vice
versa. In treating this problem, there are major computational
advantages in posing the Cauchy evolution in a frame which is
co-rotating with the orbiting black holes. Indeed, such a description
may be necessary in order to keep the numerical grid from being
intrinsically twisted and strangled. In this co-orbiting formulation of
the binary black hole problem, the Cauchy evolution requires inner
boundary conditions in two regions approximating the two disjoint
apparent horizons and also an outer boundary condition on a worldtube.
Far enough outside the outer worldtube the coordinate rotation would go
superluminal. Previous work has shown that an outgoing characteristic
code can routinely handle such superluminal gauge flows in the
exterior. Our present results indicate that an ingoing characteristic
code can just as effectively handle the inner boundaries of multiple
black holes.

Consistent worldtube data must satisfy conservation conditions
\cite{cce,highp} which correspond to a hyperbolic version of the
standard constraint equations for Cauchy data. In the present work,
since we are not matching to a Cauchy evolution, we generate this data
from an analytic solution. Specifically, we set Schwarzschild data at
the worldtube. In ingoing Kerr-Schild coordinates ${\hat x}^{\alpha}$,
the Schwarzschild metric takes the form \cite{kersch}
\begin{equation}
     ds^2 = -d{\hat t}^2 +d{\hat x}^2+d{\hat y}^2+d{\hat z}^2
            +{2m\over {\hat r}} (d{\hat t} +
            {{\hat x}d{\hat x}+{\hat y}d{\hat y}+{\hat z}d{\hat z}
             \over {\hat r}})^2,
\label{eq:kersch}
\end{equation}
where ${\hat r}^2 ={\hat x}^2 +{\hat y}^2+{\hat z}^2$ and
$k_{\mu}=-\partial_{\mu}(\hat t+\hat r)$ is the ingoing
degenerate null vector.

In order to determine data for a moving black hole, we introduce
an $x^{\alpha}$ coordinate system which is either rotating, boosted
or wobbling with respect to the ${\hat x}^{\alpha}$ coordinates. In the
$x^{\alpha}$ frame, the null coordinates are centered at $r=0$ and the
worldtube is located at $r=R$, where $r^2=x^2+y^2+z^2$. As a result, the
worldtube is stationary with respect to the null grid but the black
hole moves. 

The initial value problem also requires data on the initial null
hypersurface. In our formulation, this null data is free of
constraints, other than continuity conditions at the worldtube. Thus we
can introduce an arbitrary pulse of radiation in the data to describe a
distorted Schwarzschild black hole.  We can also pose initial null data
by setting to zero the component of the Weyl tensor intrinsic to the
null hypersurface. On a spherically symmetric null hypersurface, the
Weyl data for the Schwarzschild space-time is exactly zero. On a null
hypersurface offset from the center of spherical symmetry, this Weyl
data for a Schwarzschild spacetime is not zero (and it is not possible
to express it in simple analytical form).  In this case the choice of
vanishing Weyl data introduces an amount of gravitational radiation
which depends on the size of the offset. The details of setting up the
worldtube-nullcone data are presented in Section \ref{sec:data}.

The worldtube boundary values, as prescribed in this paper, satisfy the
conservation conditions when the spacetime is exactly Schwarzschild,
e.g. a Schwarzschild black hole in either rotating or boosted
coordinates.  But in either the distorted or wobbling case, when
gravitational radiation is contained in the initial null data, the
extraction module overdetermines the metric and its normal derivatives
at the worldtube in terms of their Schwarzschild values. As a result,
the reflection of the wave off the worldtube can lead to a violation of
the Bianchi identities. The mismatch between radiation impinging on the
Schwarzschild worldtube introduces an unphysical sheet source of
gravitational radiation on the worldtube which is not necessarily
consistent with energy and angular momentum conservation.  Although
this obscures the physical interpretation of the results for those
cases, it is remarkable that the stability of the evolution is not
affected and that the system behaves in accord with the principles of
black hole dynamics, as described in Sec. \ref{sec:results}. In a more
physical implementation, the conservation conditions would be enforced
either (i) directly, by using them to properly evolve the boundary
conditions up the worldtube, or (ii) by matching the interior evolution
across the worldtube to an exterior Cauchy evolution.

Our chief goal here is to demonstrate that the region of spacetime
interior to the world tube can be evolved {\em stably} and {\em
accurately} by means of a characteristic evolution algorithm. The
interior region is truncated in the vicinity of an apparent horizon.
The long term stability we observe indicates a surprising robustness of
the worldtube-nullcone boundary value problem.

In the case of a Schwarzschild spacetime described in a frame rotating
about the center of spherical symmetry, the location of the apparent
horizon is known analytically, as well as the transformation to null
coordinates and the null metric. Thus this case provides an important
test bed to calibrate numerical accuracy. Long term stability and
second order convergence to the analytic values have been confirmed. In
this purely rotating case in which the worldtube is a stationary
boundary, when we superimpose a pulse of radiation on the initial
Schwarzschild null data we find that the surface area of the resulting
distorted black hole grows in time but eventually reaches an
equilibrium value consistent with the Schwarzschild boundary conditions
on the worldtube. In the offset case, the Schwarzschild boundary moves
periodically but the marginally trapped surface associated with the
black hole again reaches equilibrium with it, confirming that the motion
of the boundary is ``pure gauge''.

When the null cones are not spherically symmetric a computational
approach is necessary to find the ``trapping boundary'', where a
marginally trapped surface is located on each ingoing null
hypersurface.  The analogous problem in Cauchy evolution is the
excision of a black hole interior by locating the apparent horizon.
There is an extensive literature on marginally trapped surface (MTS)
finders on spacelike hypersurfaces
\cite{nak84,cooky90,tod91,kem91,seidelsuen1992,thorn96,huq,annin,baum96,gundl}.
In Section \ref{sec:find}, we present two methods for use on null
hypersurfaces. Computational design and performance is discussed in
Sec. \ref{sec:comp}.

\section{Determination of the data}
\label{sec:data}

The worldtube data for the null evolution of the interior region is
obtained by recasting the Cartesian form of the Kerr-Schild metric Eq.
(\ref{eq:kersch}) for a Schwarzschild black hole into nonstationary
ingoing null coordinates. However, the Schwarzschild metric can only be
expressed in an analytically manageable null coordinate form when the
null cones are centered to be spherically symmetric. Thus, in the
offset case, numerical techniques must be used to carry out the
transformation from Cartesian to null coordinates in order to provide
the worldtube data.  Fortunately, the transformation need only be
implemented in the neighborhood of the world tube, This allows the
solution for the ingoing null geodesics to be carried out by means of a
Taylor expansion in affine parameter, up to the required order of
accuracy. In this way, the transformation can be formulated as a hybrid
analytical-numerical scheme, with the null properties built in
analytically.

Such a scheme for extracting null worldtube data from Cauchy data in a
general spacetime has been implemented as an {\em extraction} module
\cite{manual}, which is part of the computational procedure to match
Cauchy and characteristic evolutions across an interface. We use this
module here to obtain the required worldtube data for null evolution by
extraction from the ``3+1'' form of the Schwarzschild metric which has
been recast into nonstationary coordinates by a nontrivial choice of
lapse and shift.

The initial value problem is completed by giving data on an initial
null hypersurface. In ingoing null coordinates, with $v$ labeling the
null hypersurfaces, with $x^A =(\theta,\phi)$ labeling the angles along
the null rays and with $r$ labeling the surface area measured along the
null rays, the metric takes the Bondi-Sachs form \cite{excise}
\begin{equation}
 ds^2=\left(e^{2\beta} {V \over r} +r^2h_{AB}U^AU^B\right)dv^2
	+2e^{2\beta}dvdr -2r^2 h_{AB}U^Bdvdx^A +r^2h_{AB}dx^Adx^B,
   \label{eq:vmet} 
\end{equation}
where $det(h_{AB})= det(q_{AB})= q$, with $q_{AB}$ a unit sphere
metric.  The inverses of these 2-dimensional metrics are denoted by
$h^{AB}$ and $q^{AB}$.  We express $q_{AB}$ in terms of a complex dyad
$q_A$ (satisfying $q^Aq_A=0$, $q^A\bar q_A=2$, $q^A=q^{AB}q_B$, with
$q^{AB}q_{BC}=\delta^A_C$ and $ q_{AB} =\frac{1}{2}\left(q_A \bar
q_B+\bar q_Aq_B\right)$).  $h_{AB}$ can then be represented by its dyad
component $J=h_{AB}q^Aq^B /2$, with the spherically symmetric case
characterized by $J=0$. The full nonlinear $h_{AB}$ is uniquely
determined by $J$, since the determinant condition implies that the
remaining dyad component $K=h_{AB}q^A \bar q^B /2$ satisfies
$1=K^2-J\bar J$.  We also introduce the spin-weighted field $U=U^Aq_A$,
as well as the (complex differential) eth operators $\eth$ and $\bar
\eth$. Refer to {}~\cite{cce,eth} for further details.

The complete null data can be specified freely in terms of either the
metric quantity $J$ or the Weyl tensor component
$C_{\alpha\beta\gamma\delta}n^{\alpha} n^{\gamma} m^{\beta} m^{\delta}$
(corresponding to $\Psi_4$ in Newman-Penrose terminology \cite{np}),
where the null vector $n^{\alpha}$ and the complex spacelike vector
$m^{\alpha}$ span the tangent space to the null hypersurface. On a
spherically symmetric null hypersurface, the Weyl data for the
Schwarzschild space-time is exactly zero but since there are no
constraints on the null data, we can freely add a radiation pulse of
any desired shape.

In the case of rotating coordinates, as described in Sec.
\ref{sec:rot}, or boosted coordinates, as described in Sec.
\ref{sec:boost}, the initial null cone is spherically symmetric and
Schwarzschild null data can be determined analytically. However, it is
not possible to express Schwarzschild null data in simple analytical
form on a null hypersurface which is not spherically symmetric.
Instead, we choose initial data for an offset wobbling black hole by
setting the Weyl data to zero on the initial (nonsymmetric) null
hypersurface. The relevant components of the Riemann tensor are
\begin{eqnarray}
      R_{rABr} &=& {1\over 2} g_{AB,rr}
             -{1\over 2}g^{vr} g_{vr,r}g_{AB,r}
             -{1\over 4}g^{CD}g_{AC,r} g_{BD,r}
                  \nonumber  \\
               &=& {r^2\over 2} h_{AB,rr}
                   +r h_{AB,r}
            -r\beta_{,r}(rh_{AB,r} +2h_{AB})
             -{r^2\over 4}h^{CD}h_{AC,r} h_{BD,r}.
\end{eqnarray}
Here $g^{AB}R_{rABr}=0$ by virtue of the hypersurface equation for
$\beta$ \cite{highp},
\begin{equation}
 \beta_{,r} =- \frac{1}{16}rh^{AB}_{,r}h_{AB,r}
     = {r\over 8}\big ( J_{,r}  \bar J_{,r} -K_{,r}^2   \big ) .
\label{eq:gur}
\end{equation}
The requirement of vanishing Weyl data is equivalent to
$q^{A}q^{B}R_{rABr}=0$, which gives 
\begin{equation}
   (r^2 J_{,r})_{,r}
    -2\beta_{,r}(r^2 J)_{,r} 
    -\frac{r^2}{2} J \big (  J_{,r} \bar J_{,r}
             -K_{,r}^2 \big )
    =0 .
\label{eq:zero}
\end{equation}
Combining Eq's. (\ref{eq:gur}) and (\ref{eq:zero}), we then obtain
\begin{equation}
  r^2 (r^2 J_{,r})_{,r}
    -2\beta_{,r} (r^4 J)_{,r}  =0 .
\label{eq:init} 
\end{equation}
With $\beta_{,r}$ determined from Eq. (\ref{eq:gur}), Eq. (\ref{eq:init})
is a second order radial differential equation for the initial data
$J$, which may be solved in terms of boundary values for $J$ and
$J_{,r}$ on the worldtube.

The outer worldtube is located at $(x^2+y^2+z^2)^{1/2}=R$, in terms
of coordinates $x^{\alpha} =(t,x,y,z)$ moving with respect to static Kerr-Schild
coordinates ${\hat x}^{\alpha}$. The boundary data on the worldtube is
obtained by first transforming the metric (\ref{eq:kersch}) in ``3+1''
form to the $x^{\alpha}$ frame and then applying the {\em extraction module},
which determines numerically the transformation from a Cartesian to a
null coordinate system in the neighborhood of the worldtube. This
module supplies the boundary values of the null metric quantities $J$,
$\beta$, $U$ and $V$ on the worldtube.
 
As a check on the extraction module, we examine the rotating
case where the null Schwarzschild  metric can be found
analytically. We relate the static coordinate frame $\hat x^{\alpha}$ to the
rotating one, 
$x^{\alpha}$ by $t={\hat t}$,
$x={\hat x}\cos \omega (t + r) - {\hat y}\sin \omega (t + r)$,
$y={\hat x}\sin \omega (t + r) + {\hat y}\cos \omega (t + r)$ and
$z={\hat z}$. 
In this transformation, the angles $\theta, \phi$
(associated in the standard way with the Cartesian coordinates $x^i$) remain constant along the generators of the null cones emanating from
the world tube. Therefore, the null metric can be easily obtained by
the simple transformation $v = t+r$, $r=\sqrt{(x^2 +y^2 +z^2)}$,
$\theta = \cos^{-1}(z/r)$ and $\phi=\tan^{-1}(y/x)$, leading to
\begin{equation}
 ds^2 = \bigg( -1 + {2m\over{r}} + r^2 \omega^2 \sin^2\theta \bigg) dv^2
     + 2 dv dr - 2 r^2 \omega^2 \sin^2\theta du d\phi
     + r^2 (d\theta^2 + \sin^2\theta d\phi^2 ).
\end{equation}
This identifies the spin weighted versions of the variables appearing 
in the null metric (\ref{eq:vmet}) as
\begin{eqnarray}
J &=& 0 \: , \: \beta = 0 \\
U &=& i \, \omega \, \sin\theta \, e^{i \phi} \\
V &=& -r + 2m.
\
\label{eq:rot}
\end{eqnarray}

We used this transformation to check the accuracy of the worldtube data
extracted from the Schwarzschild metric in ``3+1'' Cauchy form with
mass $m=.25$. The code was run using the extraction radius $R = 3$.
The Cauchy grid was a Cartesian cube with  range $x^i \in [-4, 4]$ and
the range of the characteristic grid was $r \in [0, 4]$. We confirmed
convergence of the numerical error to zero at a second order rate with
grid size for several values of $\omega$ $( 0.8, 0.1, 0.001)$.

\section{Finding the trapping boundary}
\label{sec:find}

The excision of a region inside the black hole is necessary for
numerical evolution, for otherwise contact with the singular
region of spacetime would result. The boundary of the excised region
must either be trapped or marginally trapped in order to insure that it
does not contain points that can causally influence the radiated
waveform. For a slice ${\cal S}$ of an ingoing null hypersurface ${\cal
N}_v$, described in null coordinates by $r=R(v,x^A)$, the divergence of
the outgoing null normals is \cite{excise}
\begin{equation}
  {r^2 e^{2\beta}\over 2}\Theta_l = -V
           -{1\over \sqrt{q}}[\sqrt{q}(e^{2\beta}h^{AB}R_{,B}-r^2 U^A)]_{,A}
                -r(r^{-1}e^{2\beta}h^{AB})_{,r}R_{,A}R_{,B}
                +r^2 U^A_{,r}R_{,A}.
    \label{eq:diverg}
\end{equation} 
This is to be evaluated on ${\cal S}$ after all partial derivatives are
taken; e.g. $r_{,A}=0$. The slice will be marginally trapped if $\Theta_l=0$. 

Finding a marginally trapped slice on a converging ingoing null
hypersurface is a 2 dimensional elliptic problem, which entails setting
the right hand side of equation (\ref{eq:diverg}) to $0$ and solving
for $R(v,x^A)$, at a fixed advanced time $v$. It is easier to find
trapped surfaces. In fact, the largest $r=const$ slice of ${\cal N}_v$
that satisfies the algebraic inequality $Q\le 0$, where
\begin{equation}
   Q =-V+{r^2\over \sqrt{q}}(\sqrt{q} U^A)_{,A}.
    \label{eq:qmarg} 
\end{equation}
is either trapped or marginally trapped \cite{excise}. We call this
slice the ``$Q$-boundary''. A comparison of Eq's.  (\ref{eq:diverg})
and (\ref{eq:qmarg}) shows that the $Q$-boundary is marginally trapped
when the slice $Q=0$ is an $r=const$ slice. However, the gauge freedom
in the choice of a surface area coordinate ($r$ is a scalar density)
allows any slice to be regauged as an $r=const$ slice. So there is a
gauge in which the $Q$-boundary and the trapping boundary coincide. But
finding this gauge is tantamount to solving the elliptic problem for a
marginally trapped slice.

This presents us with two possible strategies for locating the inner
boundary, both of which ensure that the excised portion of spacetime
cannot causally effect the exterior spacetime: (I) Use the trapping
boundary or (II) use the Q-boundary. Strategy (I) makes the most
efficient use of the spacetime points but a 2D elliptic equation must be
solved.  Strategy (II) involves only some simple algebra so it is very
efficient computationally. We will pursue both strategies here and
compare their merits.

In implementing strategy (I), we have tried two methods for finding the
trapping boundary. One is a standard approach to the solution of the
elliptic equation $\Theta_l=0$ by solving the parabolic equation
\begin{equation}
      \partial_{\lambda} R = -\Theta_l 
     \label{eq:parab}
\end{equation}
in terms of a relaxation parameter $\lambda$. At large $\lambda$, the
solution relaxes to the location of the trapping boundary provided that
the procedure is stable.

Another method of finding the trapping boundary is by a minimization
procedure. In the case of a Cauchy hypersurface, this approach was
introduced in \cite{annin}and further developed in \cite{baum96} using
spectral methods. Here we use a finite difference version of the
minimization approach and combine it with an approach based upon
gradient flows proposed in \cite{tod91}. We combine these approaches by
characterizing the trapping boundary on a converging ingoing null cone
${\cal N}_v$ as a marginally trapped slice ${\cal S}_m$ which minimizes
the functional
\begin{equation}
 {\cal T}({\cal S}) = {1\over 8}\oint_{{\cal S}} \Theta_n^2 \Theta_l^2 dS
\end{equation} 
where $\Theta_n$ is the divergence of the ingoing null vector
$n^{\alpha}$ tangent to ${\cal N}_v$. We normalize $n^a$ and $l^a$
(both normal to ${\cal S}$) by $n^{\alpha}l_{\alpha}=-1$. With this
setup, ${\cal T}$ is invariant under changes in the extensions of the
the null normals (local boosts) that retain $n^a l_a=-1$. For a
standard sphere of radius $R$ in Minkowski space, $ {\cal T} =8\pi/R^2$
so that the minimum on a Minkowski light cone occurs at infinity. This
has the advantage of biasing the search away from the caustic tip of
the light cone when looking for nontrivial minima.

We perform a variation of the form $\delta x^{\alpha} = F n^{\alpha}
\delta \lambda$, which deforms the slice along the generators of ${\cal
N}_v$. In order to find a flow $F$ which leads toward the minimum
consider the variation
\begin{equation}
\delta {\cal T}({\cal S}) = \oint_{{\cal S}}T_{\alpha} \delta x^{\alpha} dS,
   \label{eq:calt}
\end{equation}
which serves to define $T_{\alpha}$. We choose $F=-T_{\alpha} n^{\alpha}$. Then
\begin{equation}
 \delta {\cal T}({\cal S}) = 
      -\oint_{{\cal S}}(T_{\alpha} n^{\alpha})^2\delta \lambda dS,
\end{equation}
so that the variation will lead in the direction of the minimum at 
${\cal T}=0$.

The main problem then reduces to calculating $T_{\alpha} n^{\alpha}$.
In ingoing null coordinates on ${\cal N}_v$, we describe ${\cal S}$ by
$r=R(x^A)$ and its variation by $r=R(x^A)+\delta R(x^A)$, with $\delta
v=\delta x^A =0$.  Choosing the extension
$n_{\alpha}=-g_{vr}v_{,\alpha}$, we have $\Theta_n = -2/r$ so that, for
any slice of ${\cal N}_v$, $\delta \Theta_n = 2\delta r/r^2$.

The variation of terms $f(r,x^A)$ not explicitly dependent on $R$ is
calculated using $\delta f(r,x^A) = f_{,r}\delta R$. Also, since
$dS=r^2 d\Omega$, in terms of the solid angle element in the $x^A$
coordinates, we have $\delta dS=2 dS \delta R/r$.  As a result, the
contributions from $\delta \Theta_n$ and from $\delta dS$ cancel in the
variation of Eq. (\ref{eq:calt}) so that
\begin{equation}
\delta {\cal T}({\cal S})=\oint_{{\cal S}}\Theta_l \delta\Theta_l d\Omega,
\end{equation}
From equation ({\ref{eq:diverg}),
\begin{equation}
 \delta \Theta_l =  \Theta_{l,r}\delta R      
           +{2 e^{-2\beta} \over r^2}
        \{ -{1\over \sqrt{q}}[\sqrt{q}(e^{2\beta}h^{AB}\delta R_{,B}]_{,A}
                -2r(r^{-1}e^{2\beta}h^{AB})_{,r}R_{,A}\delta R_{,B}
                +r^2 U^A_{,r}\delta R_{,A} \}
\label{eq:deltamess}
\end{equation}

For any vector field $V^A(r,x^B)$ on ${\cal S}$, we have 
\begin{equation}
    \oint_{\cal S} D_A V^A d\Omega =0,
\end{equation}
where $D_A V^A =(\sqrt{q}V^A)_{,A}/\sqrt{q} +V^A_{,r}R_{,A}$. This
allows us to eliminate terms in $\delta {\cal T}$ containing $\delta
R_{,A}$ by integrating over ${\cal S}$. We obtain
\begin{equation}
    \delta {\cal T}({\cal S}) = 
      \oint_{{\cal S}}\{\Theta_l \Theta_{l,r} +\Psi\}\delta R d\Omega
\end{equation}
where 
\begin{eqnarray}
   \Psi &=& [{2\over r^2}(\Theta_lh^{AB})_{,r}
             -{4\over r^2}(\Theta_l e^{-2\beta})_{,r}
                       e^{2\beta}h^{AB}]_{,r}R_{,A}R_{,B} 
          - [2\Theta_l e^{-2\beta}U^A_{,r}]_{,r}R_{,A}
            -[2\Theta_l e^{-2\beta}U^A_{,r}]_{,A} \nonumber \\
        &-& [{2\over r^2}(\Theta_l e^{-2\beta})_{,A}
                            e^{2\beta}h^{AB}]_{,r}R_{,B}
         +[({2\over r^2}\Theta_l h^{AB})_{,r}R_{,A}
          -({4\over r}\Theta_l e^{-2\beta})_{,r}{ e^{2\beta} \over r}
                 h^{AB}R_{,A} 
          -{2\over r^2}(\Theta_l e^{-2\beta})_{,A}e^{2\beta}h^{AB}]_{,B}.
\label{eq:Psi}
\end{eqnarray}
Thus, in order to find the trapping boundary, we follow the variational
path determined by $\delta R=-F \delta \lambda$ with
\begin{equation}
  F=(\Theta_l \Theta_{l,r} +\Psi)/R^2.
\label{eq:flow}
\end{equation}

As a check on the stability of this minimization scheme, suppose ${\cal
S}_m$ is a trapping boundary located at $r=R_m(x^A)$. Then, on ${\cal
S}_m$, $\Theta_l=0$ and $F=\Psi/R_m^2$. But the operator
$(\partial_A+R_{m,A}\partial_r)$ also annihilates $\Theta_l$ on ${\cal
S}_m$. As a result, direct substitution to eliminate the $\partial_r$
derivatives in equation (\ref{eq:Psi}) gives $\Psi =0$ on ${\cal
S}_m$.  Thus $\delta R$ also vanishes and ${\cal S}_m$ is an invariant
slice with respect to the variational scheme.

We have:

(1) A marginally trapped surface is a zero of the positive functional
${\cal T}$;
 
(2) The effect of the flow $F$ on ${\cal T}$ is everywhere negative or
zero;
   
(3) A marginally trapped surface is stationary under the flow $F$.

Thus a marginally trapped surface must be stable with respect to the
flow $F$ except in the degenerate case, corresponding to neutral
stability, where a continuum of such surfaces exist in ${\cal N}_v$.
Although this degenerate case is possible it would be improbable to
encounter in an evolution based upon a reasonably behaved foliation.
For an interesting discussion of the wild behavior possible in general
for marginally trapped surfaces see \cite{eard}.

In order to implement either of the above two finders as computational
algorithms, we represent the geometric quantities involved as
spin-weighted fields in stereographic coordinates. The spin-weighted
expressions necessary to determine $\Theta_l$ and $F$ are given in
Appendix \ref{app:a}.

\section{Computational design and performance}
\label{sec:comp}

All numerical algorithms have been based upon explicit finite
difference methods. The metric functions are discretized and placed on
a finite 3 dimensional grid, with $N_r$ radial points and $N_{\xi}^2$
angular points, whose outer boundary is a spherical worldtube. The
spherical coordinates are patched by two overlapping stereographic
grids and angular derivatives of tensors are handled by a computational
version of the $\eth$ formalism \cite{eth}.

In a general null evolution, data at the worldtube would be extracted
from a Cauchy evolution . In this example, since we are not matching to
a Cauchy evolution, we extract data at the worldtube from an
analytical Cauchy solution.  The characteristic evolution is carried
out using a code described and calibrated in \cite{highp}, transformed
into an ingoing null code according to the procedure presented in
\cite{excise}.

Inside the grid there is a black hole whose interior is partially
excised at an inner boundary, the ``hole'',  which is taken to be
either a marginally trapped surface or the Q-boundary. We need to
evolve only those grid points which are outside a discrete version of
the hole. At the same time, we also need to allow the hole to move
through the grid. In order to accomplish this we use a 3 dimensional
mask function. Each grid point is assigned the value 1 by the mask if
it either neighbors or is exterior to the boundary. All other grid
points are masked to zero.  The metric functions are evolved at each
point with mask value 1 using data only from other points with a mask
value 1 (i.e. points outside or neighboring the hole use data only from
other points outside or neighboring the hole). In case they are needed
(see below), values of the metric functions at grid points which are
nearest neighbors just inside the hole are extrapolated radially
inward using points exterior to them with mask 1. Other interior points
are ignored.

After all the metric functions have been evolved, we locate the hole at
the new time.  We then recompute the mask function and continue. If the
boundary moves out, we simply throw away the data which we just
evolved. If the boundary moves in, we have data at the new point which
was obtained by extrapolation. Using this approach saves us from having
to figure out if we have any points which were previously in the hole
but are now outside.  Such points automatically have data. It should be
noted that we can safely assume that the black hole boundary will never
move more than a grid point during any iteration. If it did, it would
violate the Courant-Friedrichs-Lewy condition, which is built into the
characteristic code to insure stability.

The procedure for locating the trapping boundary is fairly simple.
Basically, we use the previous position of the horizon ($R$) as a guess
for its current location (if this is the first iteration, we use the
position of the Q-boundary for the approximate location of the trapping
boundary).  Then, if we are finding the boundary using the parabolic
relaxation technique based upon Eq.(\ref{eq:parab}), we compute
$\Theta_l$ and then let $R = R_{old} - \Theta_l \delta \lambda$, and
repeat until $\| \Theta_l \|_2$ is less than some threshold. If
instead, we are using the minimization procedure, we compute $F$ and
then let $R = R_{old} - F \delta \lambda$, and repeat until $\| F \|_2$
is less than some threshold. The values of the stopping threshold and
$\delta \lambda$ are parameters. The threshold can be safely set to be
approximately $\Delta r$, (where $\Delta r$ is the spacing between
radial grid points) . The value of $\delta \lambda$ should be small
enough so that $F \delta \lambda < \Delta r$  but not so small that
finding the horizon requires a large number of iterations. We have
found it sufficient to choose a fixed $\delta \lambda$ at the start of
the calculation, however it is possible to design a scheme in which
$\delta \lambda$ is self-tuning and changes its value to speed
convergence of each attempt to locate the horizon.

A stability analysis of the explicit scheme (first order in time,
second order in space) used to solve Eq. (\ref{eq:parab}) shows that
$\delta \lambda$ must scale as $\Delta \xi^2$ (where $\Delta \xi$ is
the spacing between angular points). This requirement would suggest
that the proposed method is computationally expensive. However, our
results show that the use of the horizon finder introduces a negligible
overhead when dealing with long evolutions. With the aforementioned
strategy of using the position of the Q-boundary as the initial guess,
the finder may need many iterations to home in on the trapped surface,
but it takes just a few iterations thereafter to track the surface.

It is worth noting that, if instead of using the Q-boundary to
determine the initial guess one uses a more {\em educated} guess for
the location of the surface, the number of steps can be dramatically
decreased. This can be easily done in the case of the boosted black
hole (see Sec. \ref{sec:rot}). Using the expression for $r$ obtained in
that case, one can set as initial guess $R=2m/(\cos \theta \sinh \alpha
+\cosh \alpha)$, reducing the number of iterations on the first
hypersurface from several hundred (using the Q-boundary) to less
than $10$ (for values of $\alpha \leq 0.5$).

In the case of a wobbling black hole (see Sec. \ref{sec:wobbling}), we
do not know the analytic expression for the marginally trapped surface
even at the first hypersurface so that we use the Q-boundary as our
initial guess. Table \ref{wobperf} shows the number of iterations made by the horizon
finder for different values of the offset $b$ and frequency $\omega$
(for a characteristic grid having $45 \times 21^2$ points covering the
space from $r=0$ to $r=4$).  After the first ``find'' the number of
iterations necessary to track the hole is small; and since the finder
solves a $N_{\xi}^2$ problem (as opposed to an $N_r \times N_{\xi}^2$
problem), it does not add an appreciable computational time to the
overall numerical  evolution.

In comparison, we obtain decidedly inferior efficiency in locating and
tracking the hole by the minimization procedure using the flow given in
Eq. (\ref{eq:flow}). Stability analysis of the finite difference method
shows that $\delta \lambda$ must scale as $\Delta \xi^4$ in this case.
Thus, although this minimization approach is attractive, it is not
practical using finite difference techniques.  It may be possible to
improve the minimization algorithm by using pseudo-spectral techniques
\cite{nak84,gundl} to calculate the flow but we have not explored this
possibility in the present work.

\section{Results}
\label{sec:results}

Here we present some results of code runs for various initial
conditions. We describe the physical behavior of the black hole in
terms of the surface area of its marginally trapped surface.  This
surface area gives a measure of the energy of the radiation fields
introduced in the initial null data.  For the pure Schwarzschild case,
the marginally trapped surfaces have area $A_s=16\pi M_s^2$, in terms
of the mass $M_s$ of the Schwarzschild black hole. More generally, the
surface area of a marginally trapped surface determines its Hawking
mass $M_h$ \cite{hawkm} according to $A=16\pi M_h^2$. Thus, on an
ingoing null hypersurface ${\cal N}_v$, the function $\Delta(v) = M_s
-M_h(v)$ provides a measure of the energy between the marginally
trapped surface and the worldtube.

If a spacetime satisfies some suitable version of cosmic censorship,
such as asymptotic predictability, and settles down to a Kerr black
hole, then the area of any marginally trapped surface must be less than
the area of the final black hole \cite{hawkell}. In the present
context, we do not have a global asymptotically flat solution so these
results are not immediately applicable. However, if the black hole
settles down to equilibrium with the Schwarzschild boundary condition
on the worldtube, then at late advanced times we would expect
$\Delta(v)\rightarrow 0$.

\subsection{Rotating Schwarzschild black holes}
\label{sec:rot}

Our first runs are for a Schwarzschild spacetime described in
coordinates rotating about the center of spherical symmetry, as
described by the null data in Eq. (\ref{eq:rot}).  In this case, the
evolved metric is known analytically and the marginally trapped surface
is fixed at the horizon at $r=2m$, so that convergence to the exact
results can be checked.  Our results confirm that the numerically evolved
spacetime is accurate to second order in grid size. As expected, the
horizon finder converges to the known location of the  spherically 
symmetric marginally trapped surface.

\subsection{Boosted Schwarzschild black holes}
\label{sec:boost}
A boosted Schwarzschild black hole provides a test of the ability to
track the motion of a black hole and to calculate its surface area.
Let ${\hat x}^{\alpha}$ be ingoing Eddington-Finkelstein coordinates.
Define boosted coordinates $x^{\alpha}$ by ${\hat x}=x$, ${\hat y}=y$,
${\hat z}=z\cosh\alpha - t\sinh \alpha$ and 
${\hat t}=t\cosh\alpha -z\sinh\alpha$.

We locate an initial incoming null hypersurface ${\cal N}_0$ at $t=-(x^2
+y^2 +z^2)^{1/2}$. Initial Schwarzschild null data corresponds to
setting the Weyl data to zero at this initial time.  Schwarzschild
data at the extraction worldtube $x^2 +y^2 +z^2 = R^2$ is provided by
transforming the metric to the $x^{\alpha}$ coordinates.  

Let $(\hat v, \hat r, \hat \theta, \hat \phi)$ be standard ingoing null
coordinates associated with the Cartesian Eddington-Finkelstein
coordinates, and let $(v,r,\theta,\phi)$ be the null coordinates
associated with the Cartesian coordinates $x^{\alpha}$ by the
extraction module. We synchronize them so that $v=\hat v =0$ on  ${\cal
N}_0$, which is the only null hypersurface common to the $v$ and $\hat
v$ foliations. Then the boost transformation implies that $\phi =\hat
\phi$ and, on  ${\cal N}_0$, that
\begin{equation}
   \cos \hat \theta ={\cos \theta \cosh \alpha +\sinh \alpha
         \over \cosh \alpha +\cos \theta \sinh \alpha },
\end{equation}
with inverse
\begin{equation}
   \cos \theta ={\cos \hat \theta \cosh \alpha -\sinh \alpha
         \over \cosh \alpha -\cos \hat \theta \sinh \alpha },
\end{equation}
Calculation of the Jacobian of the angular transformation gives
$\hat r= r(\cos \theta \sinh \alpha +\cosh \alpha)
=r/(\cosh \alpha-\cos \hat \theta \sinh \alpha )$ on the initial
null hypersurface.

In order to find the initial null data in the boosted null frame we must
also relate $v$ and $\hat v$. Near ${\cal N}_0$, we set
$v=(t+r)+O({\hat v}^2)$. Then by carrying out the transformation to
leading order in $\hat v$ we obtain $v=\hat v/(\cosh\alpha
-\sinh\alpha\cos\hat \theta)$. This is enough to determine that
initially $J=0$ and $\beta =0$.

To the next order, the null condition
$g^{\alpha\beta}v_{,\alpha}v_{,\beta}=0$ implies $v=\hat v/(\cosh\alpha
-\sinh\alpha\cos\hat \theta)+ \kappa {\hat v}^2 +O({\hat v}^3)$ where
\begin{equation}
     \kappa_{,\hat r}=-{\sin^2 \hat \theta\sinh^2 \alpha \over
  2{\hat r}^2(\cosh \alpha -\sinh \alpha \cos\hat \theta)^3}.
\label{eq:kappar}
\end{equation}
The extraction routine is based upon the gauge condition that
$v=t$ on the worldtube, so that $v_{,t}=1$ at $r=R$. This fixes
the integration constant in Eq. (\ref{eq:kappar}) and gives
\begin{equation}
     \kappa={\sin^2 \hat \theta\sinh^2 \alpha \over
  2 \hat r(\cosh \alpha -\sinh \alpha \cos\hat \theta)^3}.
\label{eq:kappa}
\end{equation}

Similarly, the condition $g^{ab}v_{,a}\theta_{,b}=0$ that $\theta$ be
constant along the null rays implies 
\begin{equation}
   \cos \theta ={\cos \hat \theta \cosh \alpha -\sinh \alpha
         \over \cosh \alpha -\cos \hat \theta \sinh \alpha }
    +\gamma \hat v +O({\hat v}^2), 
\label{eq:cos}
\end{equation}
where
\begin{equation}
 \gamma_{,\hat r}=-{\sinh \alpha \sin^2 \hat \theta 
        \over
  {\hat r}^2(\cosh \alpha -\sinh \alpha \cos\hat \theta)^3}.
\label{eq:gammar}
\end{equation}
Here the gauge condition built into the extraction routine is
that $\theta_{,t}=0$ on the worldtube. With this boundary condition,
the integral of Eq. (\ref{eq:gammar}) gives
\begin{equation}
 \gamma={\sinh \alpha \sin^2 \hat \theta 
        \over
  {\hat r}(\cosh \alpha -\sinh \alpha \cos\hat \theta)^3}.
\label{eq:gamma}
\end{equation}

The $\hat v$ dependence of $r$ can now be obtained from
the defining equation of a surface area coordinate 
$r^4 q=\det(g_{AB})$, where $q$ is the determinant of the
unit sphere metric corresponding to the $x^A$ coordinates.
This gives $(r^4 q)_{,\hat v}=-r^4 qg_{AB}{g^{AB}}_{,\hat v}$,
where
\begin{equation}
 {g^{AB}}_{,\hat v}=2{x^A}_{,\hat a}{x^B}_{,\hat b \hat v}g^{\hat a \hat b}.
\end{equation}
A straightforward calculation on the initial null cone $v=0$ gives
\begin{equation}
     r_{,\hat v} =-{r^3 \over 2 \sin^2 \theta}
                (\gamma \gamma_{,\hat r} 
      +{\hat r}^{-2}(\cos \theta)_{,\hat \theta}\gamma_{,\hat \theta}),
\end{equation}
which, using Eq's. (\ref{eq:cos}) and (\ref{eq:gamma}), reduces to 
\begin{equation}
   r_{,\hat v} = \sinh \alpha \cos \theta.
\end{equation}

This determines the Jacobian of the transformation between the the
stationary and boosted null frames at $v=0$.  Carrying out the
transformation of the metric gives the initial null data for a boosted
Schwarzschild black hole: $J=\beta=U=0$ and
$V=-r+2m(\cosh\alpha+\sinh\alpha\cos\theta)^{-3}$, where $m$ is the
Schwarzschild mass.

After the hole has moved so that it is no longer centered about the
vertex of the null cones, the null metric still has some simple
properties at the poles ($\theta=0$ and $\theta=\pi$) due to the
axisymmetry of the system; e.g. $J=0$ at the poles. This allows an
analytic transformation between null and Kerr-Schild coordinates along
the ingoing polar null geodesics $\pm z=-(t-T) +R$, which lie on the
null foliation and leave the worldtube at $t=T$.  Along these polar
rays, $v=T+R$ and the radial null coordinate is given by $r=|z|=-(t-T)
+R$. This allows us to reexpress the location of the poles of the
horizon $\pm \hat z =2m$ analytically in null coordinates as
\begin{equation}
      r={2m \pm v\sinh \alpha \over
               \cosh \alpha \pm \sinh \alpha}.
  \label{eq:poles}
\end{equation}
Since $v=0$ on the initial null cone, the pole of the horizon
hits the vertex of the null cone after retarded time
$v= 2m/\sinh\alpha$.

%%(see figure \ref{fig:boostconst})
As a test of the evolution and finder, the surface area of the boosted
event horizon should remain constant. We observed that this is the case
throughout the evolution, modulo the first order error introduced by
the horizon finder. We have confirmed
that the surface area converges to the value determined by the
Schwarzschild mass as the grid spacing is refined.  We have also
checked that the poles of the horizon move in accord with
(\ref{eq:poles}), so that the pole which travels inward moves slightly
faster than the pole moving outward.

The algorithm performs the evolution and tracks the motion of the horizon
stably, as long as the CFL condition is satisfied. Figure
\ref{fig:boostmove} shows a 2-D cut (at $y=0$) of the horizon
displaying the position of the hole at three different
times.~\cite{animations} 

\subsection{Distorted black holes}

Here we study the approach to equilibrium of a distorted black hole
with Schwarzschild boundary conditions on a stationary worldtube.
First consider the case where the worldtube is static. We introduce a
gravitational wave pulse, with compact support, on the first
hypersurface by

\begin{equation}
   J(v=0,r,x^A) = \left\{ \begin{array}{ll}
  \displaystyle{\lambda \left(1 - \frac{R_a}{r} \right)^4 \,
                        \left(1 - \frac{R_b}{r} \right)^4 \;
                        \sqrt{\frac{4 \pi}{2 l + 1}} \; {}_{2} Y_{l,m}}
                    & \mbox{if $r  \in [R_a,R_b]$} \\
 		    & \\
                  0 & \mbox{otherwise,}
                        \end{array}
                        \right.
\end{equation}
where ${}_{2}Y_{l,m}$ is the spin two spherical harmonic, ${R_a}=1.5$,
${R_b}=3$ and the amplitude factor $\lambda=45$.

As the evolution proceeds, the pulse gets reflected by the outer
boundary and eventually falls into the hole. Our results confirm the
expected behavior of a black hole approaching equilibrium. Figure 
\ref{fig:disteq} shows the behavior of $M_h(v)$. The surface area
increases monotonically and approaches the value $16\pi M_s^2$
determined by the Schwarzschild mass of the exterior.

We also introduced a pulse on the initial hypersurface in the case
where the worldtube rotates (thus inducing a shift of its world lines
with respect to the static Schwarzschild streamlines).  We observed
that at  any given time, this does not result in any change in the
location of the boundary. Hence, a rotating world tube does not affect
the behavior of the Hawking mass confirming that the rotation is a pure
gauge effect.

\subsection{A Wobbling Black Hole}
\label{sec:wobbling}

Beginning with the Schwarzschild metric in Kerr-Schild coordinates
$\hat x^{\alpha}$, we introduce the coordinates of an offset, rotating frame
$x^{\alpha}$ by $t={\hat t}$, $x=({\hat x}+b)\cos \omega t -{\hat y}\sin
\omega t$, $y=({\hat x}+b)\sin \omega t+{\hat y}\cos \omega t$ and
$z={\hat z}$. In this frame, we use the metric and its derivatives on
the world tube $x^2 +y^2 +z^2= R^2$ to provide the boundary values for
a characteristic initial value problem based upon a family of ingoing
null hypersurfaces. Although the Schwarzschild metric is static, the
worldtube wobbles relative to the static Killing trajectories.

The initial value problem is completed by posing null data determined
by setting the Weyl data to zero on the initial null hypersurface. On a
non-spherically symmetric null hypersurface, the Schwarzschild Weyl
data is no longer zero (and it is not possible to express it in simple
analytical form). Thus our choice of vanishing Weyl data introduces an
amount of gravitational radiation which depends on the size of the
offset.

The resulting spacetime is neither spherically symmetric nor static.
Relative to the worldtube, it describes a black hole wobbling and
emitting gravitational radiation. Relative to the static Schwarzschild
symmetry, the worldtube wobbles but the black hole still moves and
radiates. This physical picture is confirmed in Sec. \ref{sec:results} by the
behavior of the surface area of the marginally trapped surface. The
results demonstrate that the region of spacetime interior to the world
tube can be evolved {\em stably} by means of a characteristic evolution
algorithm, when this interior region is truncated in the vicinity of a
trapped region. This is illustrated in Fig. \ref{fig:stabil}, which displays
the maximum values of the norms of $J$ and $U$ over the entire grid vs. time.
After an initial transient stage, they settle into a stationary state without
any sign of instability whatsoever.

Figure \ref{fig:wobbmove} displays a $z=0$ cut of the trapped surface
at different times, showing the ability to track the movement of the
hole by the horizon finder . As the evolution proceeds, the horizon
``wobbles'' through the computational grid with period $T = 2 \pi /
\omega$. We have evolved up to $2000 M$ confirming this
behavior.~\cite{animations}

The accuracy of the numerical evolution in the region exterior to the
horizon is negligibly affected by the choice of using either the
Q-boundary or marginally trapped surface as the inner boundary.  This
is illustrated, for the wobbling case, in Fig. \ref{fig:causality},
where we plot the values of $J$ vs time at points outside the inner
boundary, as obtained by both methods. The numerical values have a
negligible difference. However, evolution with the Q-boundary is
somewhat superior with respect to performance since no elliptic solver
or other iteration procedure is required.

The area of the marginally trapped surface again approaches equilibrium
with the Schwarzschild exterior. This is illustrated in
Fig. \ref{fig:wobbarea}, where the surface monotnically increases and
approaches a constant value (which converges to $16 \pi M_s^2$ in first
order). The usefulness of the Hawking mass as a measure of energy is
supported by the observation that $\Delta$ remains positive.

\section{Conclusion}

Our results display many interesting aspects of black hole physics,
although their physical understanding is not completely clear and would
require a deeper study of the surface sources induced on the worldtube.
The most important accomplishment of this work is that characteristic
evolution is now  ready to supply both the inner and outer boundary
condition for the Cauchy evolution of black holes as soon as
Cauchy-characteristic-matching is achieved.

\begin{center}
{\bf ACKNOWLEDGMENTS}
\end{center}

\vspace{0.3cm}
This work has been supported by NSF PHY 9510895 to the University of
Pittsburgh and by the Binary Black Hole Grand Challenge Alliance, NSF
PHY/ASC 9318152 (ARPA supplemented). Computer time for this project has
been provided by the Pittsburgh Supercomputing Center under grant
PHY860023P and by the National Center for Supercomputing Applications
under grant PHY970009N to Robert Marsa. We thank R. A. Matzner
and K. P. Tod for helpful comments.

\appendix

\section{Spin-weighted expressions} \label{app:a}

For the divergence of the outgoing rays, we obtain from equation
(\ref{eq:diverg}) that
\begin{equation}
{r^2 e^{2\beta}\over 2}\Theta_l = \Re \{ -V 
               +r^2(\eth \bar U+ U_{,r}\bar\eth R)
               +r(e^{2\beta}J/r)_{,r}(\bar\eth R)^2
               -r(e^{2\beta}K/r)_{,r}(\eth R)\bar\eth R 
               -\bar\eth[e^{2\beta}(K\eth R-J\bar\eth R)]\} .
\label{eq:sthetal}
\end{equation}
For $\Psi$, given in equation (\ref{eq:Psi}), we obtain 
\begin{eqnarray}
    \Psi &=& \Re \big \{ [-{2\over r^2}(\Theta_l J)_{,r}
   +{4\over r^2}e^{2\beta}J(\Theta_le^{-2\beta})_{,r}]_{,r}(\bar\eth R)^2
+[{2\over r^2}(\Theta_l K)_{,r}
 -{4\over r^2}e^{2\beta}K(\Theta_le^{-2\beta})_{,r}]_{,r}(\eth R)\bar\eth R
     \nonumber \\
&-& 2(\Theta_l e^{-2\beta} \bar U_{,r})_{,r}\eth R
      - 2\eth (\Theta_l e^{-2\beta} \bar U_{,r})
       +[{2\over r^2}e^{2\beta}(J\bar\eth -K\eth)
             (\Theta_l e^{-2\beta})]_{,r}\bar\eth R
     \nonumber \\
&+& \eth \big [({2\over r^2}\Theta_l K)_{,r}\bar\eth R 
       -({2\over r^2}\Theta_l \bar J)_{,r}\eth R 
       -({4\over r}\Theta_l e^{-2\beta})_{,r}{e^{2\beta}\over r}
                    (K\bar \eth R-\bar J\eth R)
        +{2\over r^2}e^{2\beta}(\bar J\eth 
                      -K\bar\eth)(\Theta_l e^{-2\beta})   \big ] .    
       \big \}
\label{eq:spsi}
\end{eqnarray}

%%%%%%%%%%%%%%%%%%%%%%%%%%%%%%%%%%%%%%%%%%%%%%%%%%%%%%%%%%%%%%%%%%%%%%%%%%
\begin{table}
\caption{Performance of the Horizon Finder}
\label{wobperf}
\begin{tabular}{llddd}
$\omega$ & $b$ &
First hypersurface & Second hypersurface & After the 10th hypersurface \\ 
\tableline
$0$ & $0$ & $1300$ & $1$ & $1$ \\
$0.05$& $0.05$ & $1416$ & $5$ & $\le 4$ \\
$0.1$& $0.1$ & $1800$ & $16$ & $\le 7$ \\ 
$0.2$& $0.2$ & $1539$ & $257$ & $\le 14$ \\ 
\end{tabular}
\end{table}

%%%%%%%%%%%%%%%%%%%%%%%%%%%%%%%%%%%%%%%%%%%%%%%%%%%%%%%%%%%%%%%%%%%%%%%%%%

\begin{figure}
\centerline{\epsfxsize=6in\epsfbox{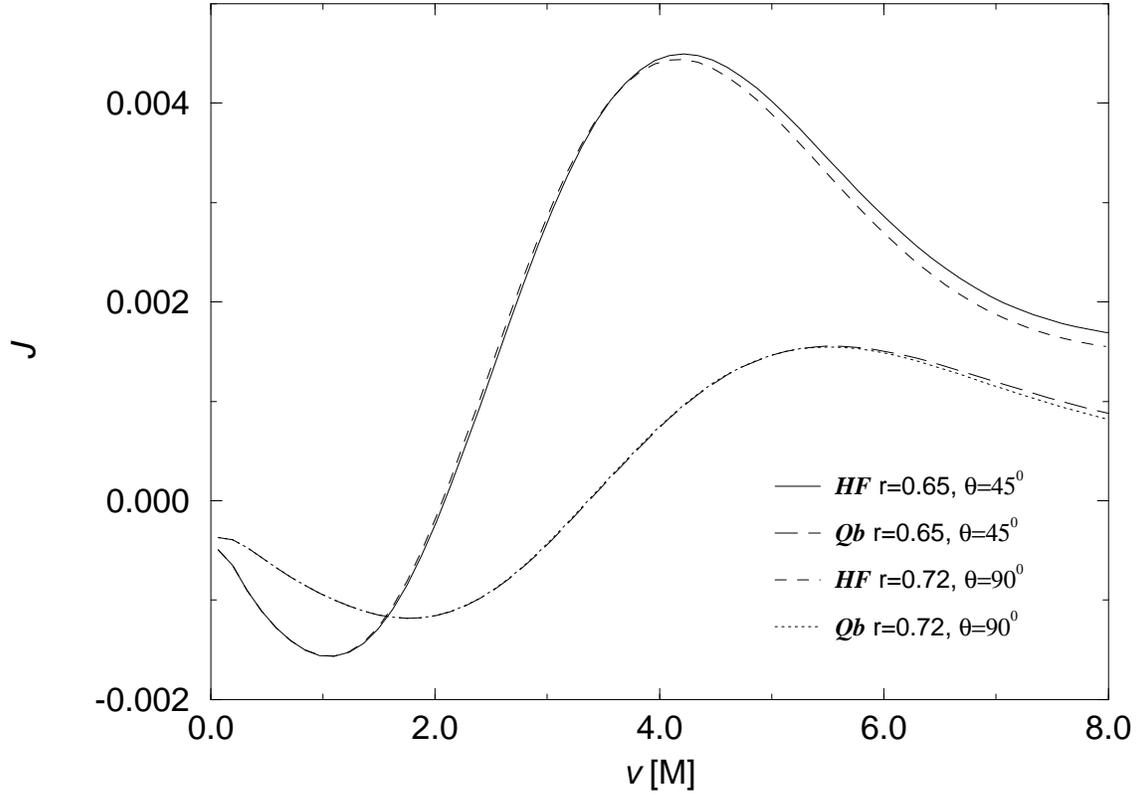}}
\caption{$J$ values at different locations obtained using the
Q-boundary (Qb) or the Horizon finder (HF); the  values obtained with the
different methods proposed to excise the hole agree
quite well, indicating not only that both methods can be applied
for the excision but also that causality is respected. The
characteristic grid had $45$ radial points and $25^2$ angular points while the Cauchy grid extended from $-4$ to $4$ with $45$ points in each direction. The
offset was $b=0.2$, the rotation frequency $\omega=0.2$ and the mass
of the Schwarzschild exterior was defined by  $m=0.5$.}
\label{fig:causality}

\end{figure}

\begin{figure}
\centerline{\epsfxsize=6in\epsfbox{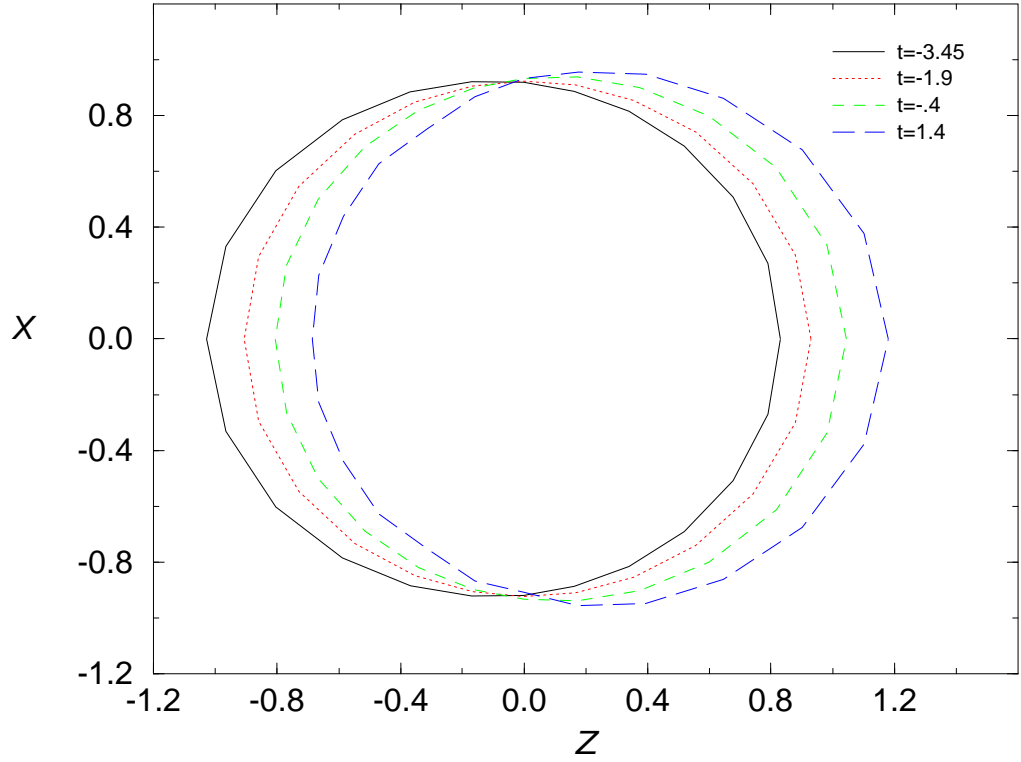}}
\caption{Tracking a boosted black hole: Cuts of the horizon at $y=0$ are
shown at times $(-3.45, -1.9, -0.4, 1.4)$.
The run was made with $41$ radial grid points and
$17^2$ angular points, with boost parameter $\alpha=0.1$ and $m=0.5$.}
\label{fig:boostmove}
\end{figure}

\begin{figure}
\centerline{\epsfxsize=6in\epsfbox{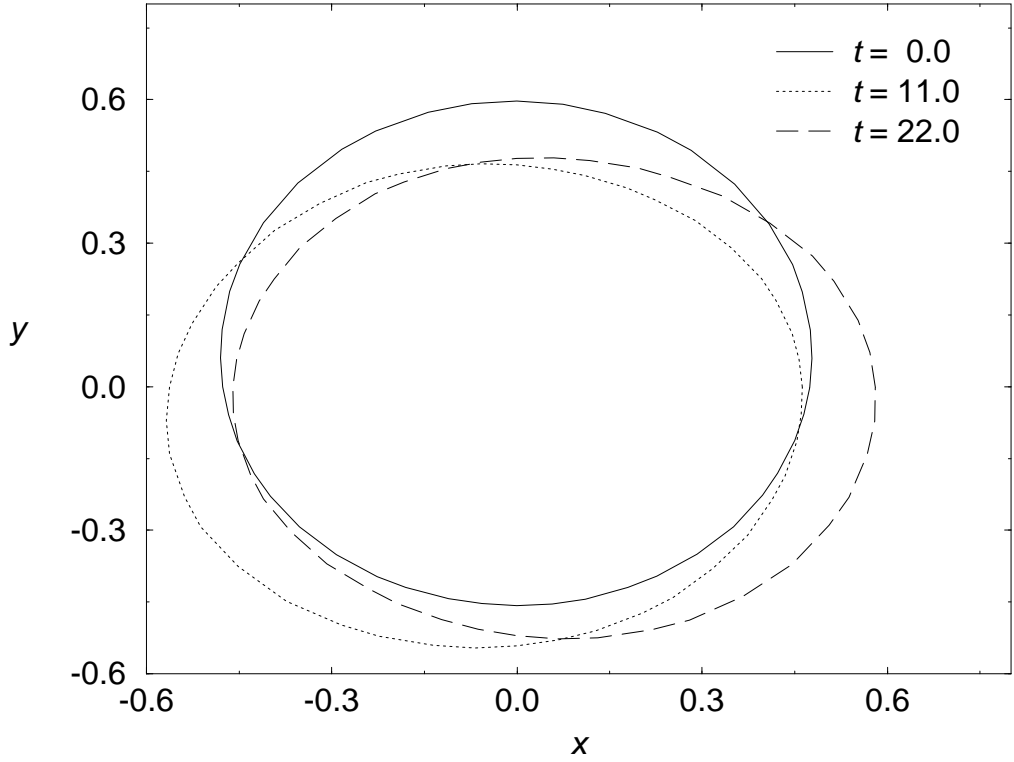}}
\caption{Tracking a wobbling black hole: Cuts of the horizon at $z=0$ at times $(0.0, 11.0, 22.0)$ displaying how the horizon finder can track the
movement of the black hole. The run was made with $45$ radial grid points and
$25^2$ angular points. We also set $m=0.25$, offset $b=0.1$ and angular velocity $\omega=0.1$.}
\label{fig:wobbmove}
\end{figure}

\begin{figure}
\centerline{\epsfxsize=6in\epsfbox{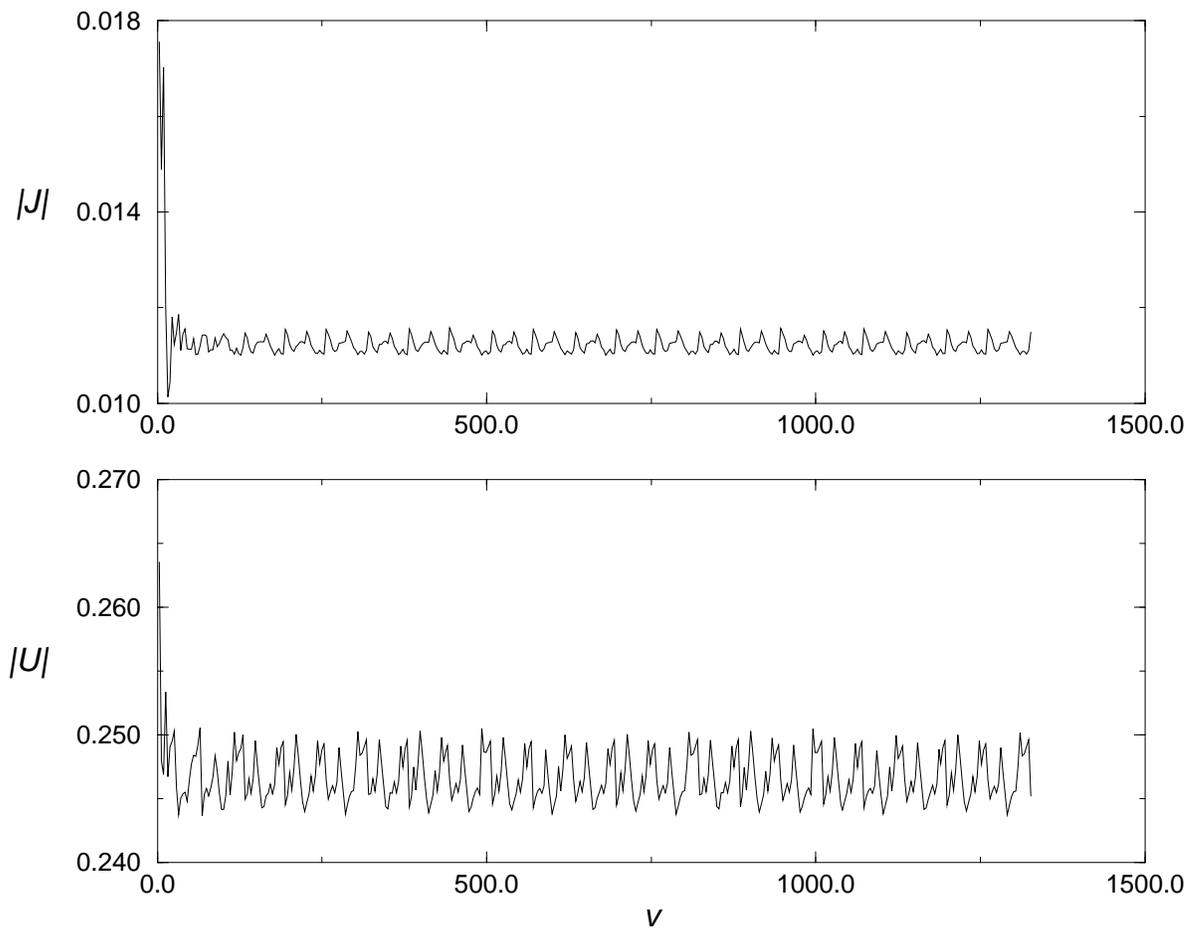}}
\caption{Stability of a wobbling black hole: Maximun values of $|J|$ and
$|U|$  over the entire grid vs. time (for a run until $v=1400 M$). After an initial stage, these values 
settle to a stationary state indicating the stability of the evolution. The run was made with $45$ radial grid points and
$25^2$ angular points. We also set $m=0.25$, offset $b=0.2$ and angular velocity $\omega=0.2$.}
\label{fig:stabil}
\end{figure}

\begin{figure}
\centerline{\epsfxsize=6in\epsfbox{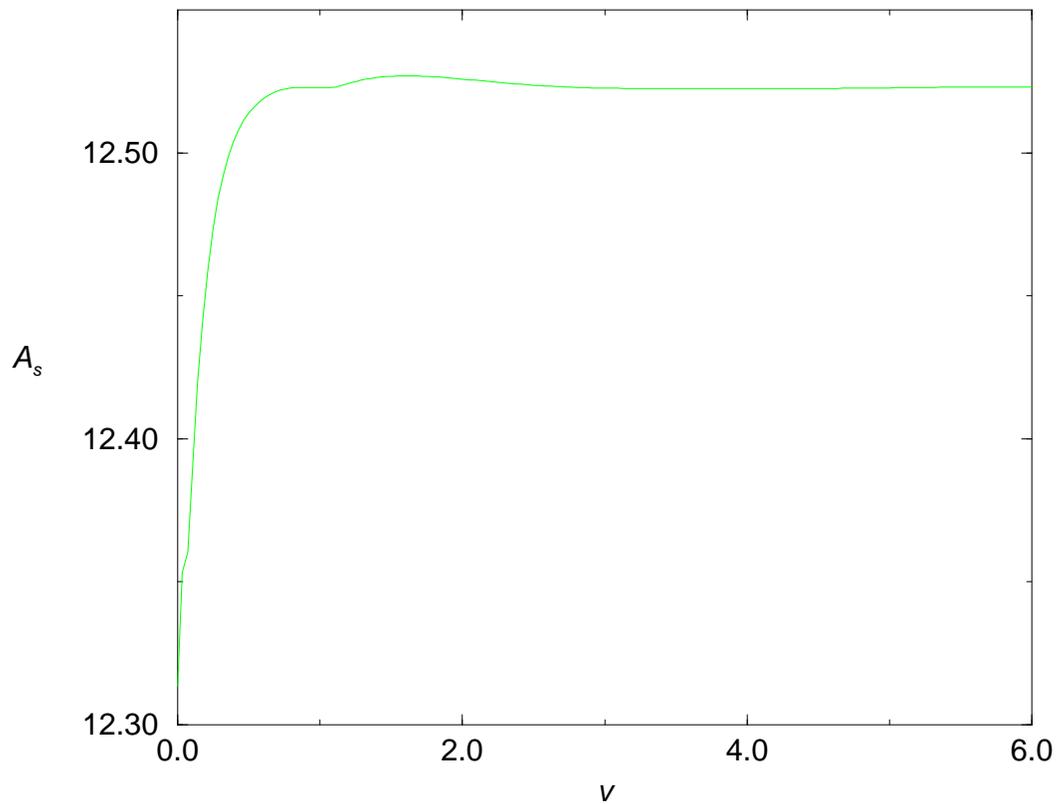}}
\caption{Behavior of the surface area vs. time for a distorted black
hole: At late time the system approaches equilibrium. The amplitude of
the pulse is $\lambda=45$ describing an $l=2$, $m=0$ spin weight $2$ pulse,
extending  from $r=1.5$ to $r=3.0$ at the first hypersurface. The code was
run with $41$ radial grid points and $17^2$ angular points, setting $m=0.5$}
\label{fig:disteq}
\end{figure}

\begin{figure}
\centerline{\epsfxsize=6in\epsfbox{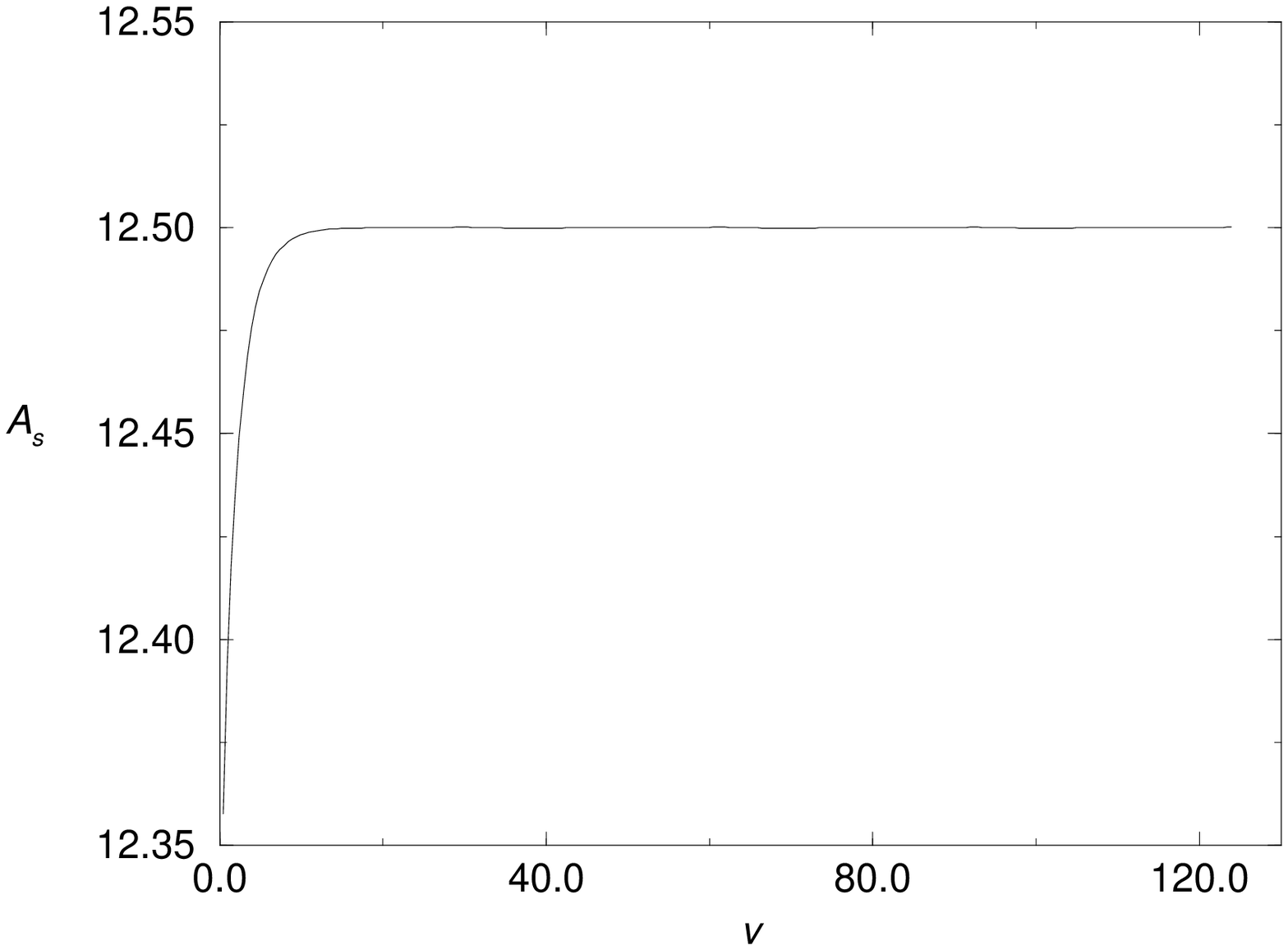}}
\caption{Behavior of the surface area vs. time for a wobbling black
hole: At late time, the area approaches a constant value. The run was
made with $45$ radial grid points and
$21^2$ angular points, with offset $b=0.1$ and angular velocity
$\omega=0.1$. The mass of the Schwarzschild exterior $m$ was set to $0.5$.}
\label{fig:wobbarea}
\end{figure}

%%%%%%%%%%%%%%%%%%%%%%%%%%%%%%%%%%%%%%%%%%%%%%%%%%%%%%%%%%%%%%%%%%%%%%%%%%


\begin{references}

\bibitem{science}R.~A. Matzner, H.~E. Seidel, S. L. Shapiro, 
L. Smarr, W-M Suen, S. A. Teukolsky, and J. Winicour, {\em Science}
{\bf 270}, 941 (1995).
\bibitem{bishop1996} N. T. Bishop, R. G\'omez, P. R. Holvorcem, R. A. Matzner,
P. Papadopoulos, and J. Winicour, Phys. Rev. Lett. {\bf 76},
4303 (1996).
\bibitem{jcp} N. T. Bishop, R. G\'omez, P. R. Holvorcem, R. A. Matzner,
P.  Papadopoulos, and J. Winicour, {\it J. Comput. Phys.} {\bf 136},
236 (1997).
\bibitem{cce} N. T. Bishop et al., Phys Rev D {\bf 54}, 6153 (1996).
\bibitem{highp}  N. T.  Bishop, R. G\'{o}mez, L. Lehner, M. Maharaj and
J. Winicour, {\it Phys. Rev. D} {\bf 56}, No. 10 (Nov. 1997),
gr-qc/9708065.
\bibitem{kersch} R. P. Kerr and A. Schild, in {\it Proceedings of the
Galileo Galilei Centenary Meeting on General Relativity, Problems of
Energy and Gravitational Waves}, ed. G Barbera (Florence:  Comitato
Nazionale per le Manifestazione Celebrative, 1965).
\bibitem{manual} N. T.
Bishop, R. G\'{o}mez, R. A. Isaacson, L. Lehner, Bela Szilagyi and J.
Winicour, ``Cauchy Characteristic Matching'', in {\it On the Black Hole Trail},
ed. B. Iyer (Kluwer, to appear).
\bibitem{thornburg1987} J. Thornburg, Class. Quantum Grav. {\bf 4}, 1119
(1987).
\bibitem{excise} R. G\'{o}mez, R. L. Marsa and J. Winicour, ``Black
hole excision with matching'', {\it Phys. Rev. D} {\bf 56}, No. 10
(Nov.  1997), gr-qc/9708002.
\bibitem{losalam} M. Huq {\em Static and boosted single hole results}. ; 
M. Scheel {\em Causal differencing overview: translating the hole}, 
in {\it Proceedings of the Conference on Astrophysical Black Holes} ed. R.A. Matzner 
(Los Alamos, 1997) (available at the website http://www.npac.syr.edu/projects/bh/).
\bibitem{nak84} T. Nakamura, Y. Kojima and K. Oohara, Phys. Lett. {\bf
106A}, 235 (1984).
\bibitem{cooky90} G. Cook and J. W. York, Phys. Rev. D {\bf 41}, 1077 (1990).
\bibitem{tod91} K. P. Tod,  Class. Quantum Grav. {\bf 8}, L115 (1991).
\bibitem{kem91} A. J. Kemball and N. T. Bishop,  Class. Quantum Grav. 
{\bf 8}, 1361 (1991).
\bibitem{seidelsuen1992} E. Seidel and W. Suen., Phys. Rev. Lett. {\bf 69},
1845 (1992).
\bibitem{thorn96} J. Thornburg, Phys. Rev. D {\bf 54}, 4899 (1996)
\bibitem{huq} M. F. Huq, S. A. Klasky, M. W. Choptuik and R. A. Matzner,
(unpublished).
\bibitem{annin} P. Anninos, K. Camarda, J. Libson, J. Mass\'{o},
E. Seidel and W. Suen, ``Finding Apparent Horizons in Dynamic 3D
Numerical Spacetimes'', gr-qc 9609059.
\bibitem{baum96} T. W. Baumgarte, G. B. Cook, M. A. Scheel, S. L. Shapiro
and S. A. Teukolsky, Phys. Rev. D {\bf 54}, 4849 (1996).
\bibitem{gundl}C. Gundlach,``Pseudo-spectral apparent horizon finders:
an efficient new algorithm'', gr-qc 9707050.
\bibitem{eard} D. M. Eardley, ``Black hole boundary conditions and
coordinate conditions'', gr-qc 9703027.
\bibitem{eth} R. G\'{o}mez, L. Lehner, P. Papadopoulos and J. Winicour,
{\em Class. Quantum Grav.}, {\bf 14} 977, 1997.
\bibitem{np} E. T. Newman and R. Penrose, J. Math. Phys. {\bf 3}, 566
(1962).
\bibitem{hawkm}S. W. Hawking, J. Math. Phys. {\bf 9}, 598 (1968).
\bibitem{hawkell} S.~W. Hawking, and  G.~F.~R. Ellis, {\em The Large
Scale Structure of Spacetime} (Cambridge University Press, Cambridge,
1973).
\bibitem{animations} Animations of these results can be viewed at the
web site http:// artemis.phyast.pitt.edu/animations


\end{references}
\end{document}